\documentclass[12pt]{article}
\usepackage{amssymb}
\usepackage{epsfig}
\usepackage{citesort}

\usepackage{graphicx}

\def\L{\mathcal L}
\def\e{\varepsilon}

\newcommand{\wt}{\widetilde}

\begin{document}

\def\a{\alpha}
\def\b{\beta}
\def\c{\chi}
\def\d{\delta}
\def\e{\epsilon}
\def\f{\phi}
\def\g{\gamma}
\def\h{\eta}
\def\i{\iota}
\def\j{\psi}
\def\k{\kappa}
\def\l{\lambda}
\def\m{\mu}
\def\n{\nu}
\def\o{\omega}
\def\p{\pi}
\def\q{\theta}
\def\r{\rho}
\def\s{\sigma}
\def\t{\tau}
\def\u{\upsilon}
\def\x{\xi}
\def\z{\zeta}
\def\D{\Delta}
\def\F{\Phi}
\def\G{\Gamma}
\def\J{\Psi}
\def\L{\Lambda}
\def\O{\Omega}
\def\P{\Pi}
\def\Q{\Theta}
\def\S{\Sigma}
\def\U{\Upsilon}
\def\X{\Xi}

\def\ve{\varepsilon}
\def\vf{\varphi}
\def\vr{\varrho}
\def\vs{\varsigma}
\def\vq{\vartheta}

\def\dg{\dagger}                                     
\def\ddg{\ddagger}                                   
\def\wt#1{\widetilde{#1}}                    
\def\mt{\widetilde{m}_1}
\def\mti{\widetilde{m}_i}
\def\rt{\widetilde{r}_1}
\def\mtt{\widetilde{m}_2}
\def\mttt{\widetilde{m}_3}
\def\rtt{\widetilde{r}_2}
\def\mb{\overline{m}}
\def\VEV#1{\left\langle #1\right\rangle}        
\def\be{\begin{equation}}
\def\ee{\end{equation}}
\def\ds{\displaystyle}
\def\ra{\rightarrow}

\def\bea{\begin{eqnarray}}
\def\eea{\end{eqnarray}}
\def\NO{\nonumber}
\def\Bar#1{\overline{#1}}


\def\pl#1#2#3{Phys.~Lett.~{\bf B {#1}} ({#2}) #3}
\def\np#1#2#3{Nucl.~Phys.~{\bf B {#1}} ({#2}) #3}
\def\prl#1#2#3{Phys.~Rev.~Lett.~{\bf #1} ({#2}) #3}
\def\pr#1#2#3{Phys.~Rev.~{\bf D {#1}} ({#2}) #3}
\def\zp#1#2#3{Z.~Phys.~{\bf C {#1}} ({#2}) #3}
\def\cqg#1#2#3{Class.~and Quantum Grav.~{\bf {#1}} ({#2}) #3}
\def\cmp#1#2#3{Commun.~Math.~Phys.~{\bf {#1}} ({#2}) #3}
\def\jmp#1#2#3{J.~Math.~Phys.~{\bf {#1}} ({#2}) #3}
\def\ap#1#2#3{Ann.~of Phys.~{\bf {#1}} ({#2}) #3}
\def\prep#1#2#3{Phys.~Rep.~{\bf {#1}C} ({#2}) #3}
\def\ptp#1#2#3{Progr.~Theor.~Phys.~{\bf {#1}} ({#2}) #3}
\def\ijmp#1#2#3{Int.~J.~Mod.~Phys.~{\bf A {#1}} ({#2}) #3}
\def\mpl#1#2#3{Mod.~Phys.~Lett.~{\bf A {#1}} ({#2}) #3}
\def\nc#1#2#3{Nuovo Cim.~{\bf {#1}} ({#2}) #3}
\def\ibid#1#2#3{{\it ibid.}~{\bf {#1}} ({#2}) #3}

\title{\bf Leptogenesis with an almost conserved lepton number}
\author{Takehiko Asaka$^a$ and Steve Blanchet$^b$\\
$^a$\emph{{\small Department of Physics, Niigata University, 950-2181 Niigata,
 Japan}}\\
$^b$\emph{{\small Maryland Center for Fundamental Physics,
University of Maryland, College Park, MD 20742, USA}}}

\maketitle

\begin{abstract}
Seesaw models with a slightly broken lepton number symmetry can
explain small neutrino masses, and allow for low-scale leptogenesis.
We make a thorough analysis of leptogenesis within the simplest
model with two right-handed (RH) neutrinos (or with $N_3$
decoupled). We obtain a semi-analytical formula for the final
asymmetry in both supersymmetric and non-supersymmetric cases with a
simple dependence on each parameter. The low-energy parameters
factorize from the high-energy ones, and the high-energy phase must
be non-zero. The role of the PMNS phases is carefully studied.
Moreover, we find that the breaking parameter in the Yukawa coupling
matrix must be relatively large, $\e_h\gtrsim 10^{-3}$ for normal
and $10^{-2}$ for inverted hierarchy. Therefore, leptogenesis in our
simple model is incompatible with RH neutrino signals at future
colliders or sizable lepton-flavor violation. The other breaking
parameter, $\e_M$, which appears in the RH neutrino mass matrix, can
be much smaller, and actually needs to be so in order to have
low-scale leptogenesis.
\end{abstract}

\pagebreak

\section{Introduction}

Leptogenesis \cite{Fukugita:1986hr} is one of the most attractive
scenarios to explain the origin of the observed matter-antimatter
asymmetry of the Universe. It follows from the seesaw mechanism
\cite{Minkowski:1977sc,Yanagida,Gell-Mann,Glashow,Barbieri:1979ag,Mohapatra:1980yp},
which gives a natural and simple explanation to the small neutrino
masses observed in neutrino experiments, and relies on the
conversion of a lepton asymmetry to a baryon asymmetry thanks to the
non-perturbative sphaleron processes \cite{Kuzmin:1985mm}. For a
recent review on leptogenesis, see~\cite{Davidson:2008bu}.

If neutrinos are massive Majorana particles, then lepton number must
be violated. Since neutrino masses are observed experimentally to be
tiny, a slightly broken lepton number symmetry, e.g. a global
$U(1)_L$, could provide the explanation. In this case, small
neutrino masses are not explained by a `seesaw' mechanism, but
rather by a cancellation
mechanism~\cite{Wyler:1982dd,Bernabeu:1987gr,Branco:1988ex}. Such a
symmetry was introduced in the context of the $\n$MSM
\cite{Asaka:2005an} in \cite{Shaposhnikov:2006nn} to explain at the
same time the keV scale of the dark matter sterile
neutrino~\cite{Asaka:2005an} and the quasi-degeneracy of the heavier
two RH neutrinos, supposed to explain the baryon asymmetry of the
Universe by means of leptogenesis via neutrino oscillations
\cite{Asaka:2005pn,Akhmedov:1998qx}. The work~\cite{Kersten:2007vk}
also made use of a slightly broken lepton symmetry to motivate large
Yukawa couplings with TeV masses for the RH neutrinos, making them
in principle accessible at the LHC.

One consequence of the slightly broken $U(1)_L$ symmetry is the
existence of two quasi-degenerate RH neutrinos. This is interesting
in the context of leptogenesis because it allows for an enhancement
of the $C\!P$ asymmetry
parameter~\cite{Covi:1996wh,Pilaftsis:1997jf}, and hence successful
leptogenesis is possible much below the usually quoted bounds on the
mass scale and on the reheat temperature of the Universe after
inflation assuming hierarchical RH neutrinos, $M\,(T_{\rm
reh})\gtrsim 3\,(1.5)\times 10^9~{\rm GeV}$
\cite{Davidson:2002qv,Buchmuller:2002rq,Blanchet:2006be,Antusch:2006gy}.
The tension with the gravitino overproduction in mSUGRA scenarios
\cite{Khlopov:1984pf,Ellis:1995mr,Moroi:1993mb,Pradler:2006hh} is
thus relaxed. The second consequence of the broken symmetry is the
presence of large washout parameters, implying that the asymmetry
will be completely independent of the initial conditions, i.e. both
the initial number of RH neutrinos and any previously generated
asymmetry, even taking into account flavor
effects~\cite{Nardi:2006fx,Abada:2006fw}.

In this paper, we study in detail the mechanism of leptogenesis in
the presence of a slightly broken $U(1)_L$ symmetry within the
supersymmetric (SUSY) and non-supersymmetric two RH neutrino (2RHN)
model~\cite{Frampton:2002qc,Ibarra:2003xp,Ibarra:2003up}. Note that
the 2RHN model is physically equivalent to the $N_3$ decoupling
limit ($M_3\to \infty$)~\cite{Ibarra:2003xp,Chankowski:2003rr}. A
related study was performed in~\cite{Pilaftsis:2005rv} with three
quasi-degenerate RH neutrinos, which have to be motivated by a
larger symmetry group, e.g. $SO(3)$. There, the focus was to find
numerical examples where resonant leptogenesis was possible with at
the same time phenomenological consequences like observable
lepton-flavor-violating signals in the non-SUSY setup. Here, we
introduce only a $U(1)_L$ symmetry, and tackle the problem with only
two RH neutrinos in both SUSY and non-SUSY cases. Proceeding
analytically, we keep under control the full parameter space of the
problem, and we do not focus on resonant leptogenesis, which
corresponds to the maximal possible enhancement of the $C\!P$
asymmetry for quasi-degenerate heavy neutrinos.

The 2RHN model implies one massless light neutrino and a reduction
of the numbers of parameters compared to the model with three RH
neutrinos from 18 to 11, among which 7 (2 neutrino masses, 3 mixing
angles and 2 $C\!P$-violating phases) are accessible in experiments.
The lower number of parameters will allow us to have a perfect
handle on the problem, and thus we will be able to derive an
expression for the baryon asymmetry predicted by leptogenesis where
the dependence on each parameter is simple. In particular, the
high-energy parameters will factorize from the low-energy ones and
from flavor effects altogether. This will make possible to study in
detail the dependence of the predicted baryon asymmetry on the Dirac
and Majorana $C\!P$-violating phases as well as the unknown angle
$\q_{13}$. Note that, since the high-energy phase will be required
to be non-zero for successful leptogenesis, leptogenesis from
exclusively low-energy $C\!P$
violation~\cite{Blanchet:2006be,Pascoli:2006ie,Pascoli:2006ci,Branco:2006ce,Molinaro:2007uv,Anisimov:2007mw}
is not viable here. Finally, we will use the maximal enhancement of
the $C\!P$ asymmetry in the resonant
limit~\cite{Pilaftsis:1997jf,Pilaftsis:2003gt,Anisimov:2005hr} to
obtain a constraint on one of the breaking parameters.
Interestingly, this constraint implies that successful leptogenesis
is incompatible with the possible observation of RH neutrinos at
future
colliders~\cite{delAguila:2005mf,Han:2006ip,delAguila:2007em,Kersten:2007vk}
as well as sizable lepton-flavor-violating signals. This conclusion
holds also in the SUSY case, if we assume the scale of leptogenesis
to be below $10^6$~GeV to avoid the gravitino problem.

In Section 2, we introduce the parametrization of the 2RHN model
which we will use throughout the paper. In Section 3, we describe
how the lepton number symmetry affects the structure of the neutrino
Yukawa matrix and the Majorana mass matrix. Then, we discuss how to
parametrize the small breaking of this symmetry. In Section 4, we
turn to leptogenesis, and estimate the baryon asymmetry predicted in
this model with a special emphasis on the role of the PMNS phases.
We also constrain the size of the breaking parameters using the
maximal enhancement of the $C\!P$ asymmetry. In Section 5, we extend
the results to the supersymmetric version of the model. Finally, we
summarize our main results and conclude in Section 6.

\section{Parametrization}

Consider the seesaw model with two RH neutrinos. In the basis where
the charged leptons and RH Majorana neutrino mass matrices are both
diagonal, the ``mass'' basis, the seesaw mass matrix is given by \be
M_{\n}=-M_D M_M^{-1}M_D^T, \ee where \be M_M\equiv D_N={\rm
diag}(M_1,M_2). \ee This matrix can be diagonalized as \be
U^{\dagger}M_{\n}U^{\star}\equiv D_{\n}={\rm diag}(m_1,m_2,m_3), \ee
where $U$ is the PMNS matrix. We will adopt the
parametrization~\cite{PDBook}
\begin{equation}\label{Umatrix}
U=\left( \begin{array}{ccc}
c_{12}\,c_{13} & s_{12}\,c_{13} & s_{13}\,{\rm e}^{-{\rm i}\,\d} \\
-s_{12}\,c_{23}-c_{12}\,s_{23}\,s_{13}\,{\rm e}^{{\rm i}\,\d} &
c_{12}\,c_{23}-s_{12}\,s_{23}\,s_{13}\,{\rm e}^{{\rm i}\,\d} & s_{23}\,c_{13} \\
s_{12}\,s_{23}-c_{12}\,c_{23}\,s_{13}\,{\rm e}^{{\rm i}\,\d}
& -c_{12}\,s_{23}-s_{12}\,c_{23}\,s_{13}\,{\rm e}^{{\rm i}\,\d}  &
c_{23}\,c_{13}
\end{array}\right)
\times {\rm diag(1, {\rm e}^{{\rm i}\,{\phi\over 2}}, 1)}
\, ,
\end{equation}
where $s_{ij}\equiv \sin\theta_{ij}$, $c_{ij}\equiv\cos\theta_{ij}$
and, neglecting statistical errors, we will use $\theta_{12}=\pi/5$
and $\theta_{23}=\pi/4$, compatible with the results from neutrino
oscillation experiments. Moreover, we will adopt the $3\s$ range
$s_{13}=0\textrm{--}0.20$~\cite{GonzalezGarcia:2007ib}.

Neutrino oscillation experiments measure two neutrino mass-squared
differences, $\Delta m_{\rm sol}^2 \simeq 8 \times 10^{-5}$ eV$^{2}$
and $\Delta m_{\rm atm}^2 \simeq 2.5 \times 10^{-3}$ eV$^{2}$~\cite{GonzalezGarcia:2007ib}.
Recall that with only two RH neutrinos the lightest active neutrino is massless.
Therefore, if the neutrino mass hierarchy is normal, one has
$m_1=0$, $m_2=\sqrt{\Delta m_{\rm sol}^2}$
and $m_3=\sqrt{\Delta m_{\rm sol}^2 + \Delta m_{\rm atm}^2}$, whereas if it is inverted,
 $m_3=0$, $m_2=\sqrt{\Delta m_{\rm atm}^2}$ and
$m_1=\sqrt{\Delta m_{\rm atm}^2 - \Delta m_{\rm sol}^2}$.

Let us now introduce the so-called Casas-Ibarra parametrization
\cite{Casas:2001sr}
\begin{equation}\label{h} M_D=U D_{\n}^{1/2}\O
D_{N}^{1/2}.
\end{equation}
The $\O$ matrix is a $3\times 2$ matrix which can be parametrized
as~\cite{Ibarra:2003xp,Petcov:2005yh}
\begin{equation}\label{Omega}
\Omega^{\rm NH}=\left(
\begin{array}{cc}
  0  &  0   \\
  \pm \sqrt{1-{\omega}^2} & - {\omega} \\
  \xi \, {\omega} & \pm \xi\, \sqrt{1-{\omega}^2}
\end{array}
\right) ,\quad \Omega^{\rm IH}=\left(
\begin{array}{cc}
    \pm \sqrt{1-{\omega}^2 } & - {\omega}  \\
  \xi\, {\omega}  & \pm \xi\, \sqrt{1-{\omega}^2 }\\
0&0
\end{array}
\right),
\end{equation}
in the normal and inverted hierarchy, respectively, and where $\omega$ is a complex
parameter. $\xi=\pm 1$ is a discrete parameter that accounts for a discrete
indeterminacy in $\Omega$.

\section{Lepton number symmetry}

Following~\cite{Shaposhnikov:2006nn} (see
also~\cite{Branco:1988ex,Kersten:2007vk}) in the limit
of only two RH neutrinos, we introduce a global $U(1)_L$ lepton number
symmetry for the leptonic fields, lepton doublets $\ell_{\a}~(\a=e,\m,\t)$
and right-handed neutrinos $\tilde{N_i}~(i=1,2)$. The $U(1)_L$ charge assignments
are as follows:
\begin{equation}
Q(\ell_{e})=+1,~Q(\ell_{\m})=+1,~Q(\ell_{\t})=+1,~Q(\tilde{N}_1)=+1, ~Q(\tilde{N}_2)=-1.
\end{equation}
The charges of all the other Standard Model fields are zero. Note that we work
in the basis where the mass matrix of the charged leptons is real and
diagonal. From now on, we shall call the basis with $\tilde{N}_i$ the
``flavor'' basis. In this basis, the relevant terms in the Lagrangian are
\begin{equation}\label{lagrangian}
\mathcal{L}= \mathcal{L}_{\rm SM} - \tilde{h}_{\a i} \bar{\ell}_{\a}
P_R \tilde{N}_i {\rm i} \s_2\F^* - {1\over 2}[\tilde{M}_{M}]_{ij}
\overline{\tilde{N}_i^c}\tilde{N}_j +h.c.\quad (i=1,2,\quad
\a=e,\m,\t),
\end{equation}
where $\tilde{h}$ and $\tilde{M}_M$ are the Yukawa coupling matrix
and the Majorana mass matrix in the flavor basis, respectively.

First of all, let us consider the $U(1)_L$ symmetry to be exact. In this case,
the allowed structure of $\tilde{h}$ is given by
\begin{equation}
\tilde{h}_0=\left(\begin{array}{ll}
\tilde{h}_{e1}&0\\
\tilde{h}_{\m1}&0\\
\tilde{h}_{\t1}&0\end{array}\right)\equiv \left(\tilde{h}_{\a1},0\right).
\end{equation}
Without loss of generality, by using the field redefinition of $\ell_{\a}$, the
non-zero components $\tilde{h}_{\a1}$ can be taken real and positive.

On the other hand, the allowed form of $\tilde{M}$ is
\begin{equation}
\tilde{M}_{M,0}=\left(\begin{array}{cc}
0&M\\
M&0\end{array}\right),
\end{equation}
where the mass parameter $M$ can be taken real and positive by field
redefinitions of $\tilde{N}_1$ and $\tilde{N}_2$.

It should be noted that, when $U(1)_L$ is exact, the masses of the active
neutrinos are exactly zero. This is because lepton doublets carry
non-zero charges, and hence a Majorana mass term for them is forbidden.

Let us now consider the effects of $U(1)_L$ symmetry breaking. Such
effects induce additional contributions to $\tilde{h}$ and $\tilde{M}$:
\begin{eqnarray}
\tilde{h}&=&\tilde{h}_0+\tilde{h}_1=h\left(\tilde{f}_{\a1},
\e_h\tilde{f}_{\a2}\right)\label{tildeF}\\
\tilde{M}_M &=&\tilde{M}_{M,0}+\tilde{M}_{M,1}=\left(\begin{array}{cc}
\e_M m_{11}&M\\
M&\e_M m_{22}e^{{\rm i}\b}\end{array}\right),
\end{eqnarray}
where $\tilde{f}_{\a i}$ are complex parameters with modulus of
order one, $m_{ii}$ is real and positive, $\b$ is an arbitrary
phase, and $\e_M$ and $\e_h$ are breaking parameters, which are
supposed to be much smaller than one. Note that we will not deal
here with the issue of the $U(1)_L$ breaking mechanism, and hence
just parametrize the breaking parameters as given.

We go now from the flavor basis to the mass basis for the RH
neutrinos. The Majorana mass matrix $\tilde{M}_M$ is symmetric, and
therefore can be diagonalized using a unitary matrix $U_N$:
\begin{equation}
U_N^T\tilde{M}_M U_N =M_M={\rm diag}(M_1,M_2).
\end{equation}
Thus, the flavor eigenstates $\tilde{N}_i$ are related to the mass
eigenstates $N_i$ by $\tilde{N}_i=[U_N]_{ij} N_j$.
It is easy to show that, to first order in $\e_M$, the eigenvalues
of $\tilde{M}_M$ are given by
\begin{equation}
M_1=M-{m_0\over 2} \e_M,~M_2=M+{m_0\over 2}\e_M,
\end{equation}
where $m_0=\sqrt{m_{11}^2+m_{22}^2+2m_{11}m_{22}\cos\b}$. So the
degeneracy parameter $\d_2=(M_2-M_1)/M_1$ is equal to the breaking
parameter $\e_M$ up to a factor $m_0/M$ supposed to be of order one.
For simplicity we will assume in the remainder of the paper that
$m_0/M=1$.

Next, we turn to the neutrino Yukawa matrix $\tilde{h}$, which we
want to express in the mass basis, i.e. $h=\tilde{h}U_N$. In a first
approximation, we take $\e_M\to 0$, so that the mixing matrix of
$N_1$ and $N_2$ is simply given by
\begin{equation}\label{V}
U_N={1\over \sqrt{2}}\left(\begin{array}{cc}
1&1\\
-1&1\end{array}\right)\left(\begin{array}{cc}
{\rm i}&0\\
0&1\end{array}\right),
\end{equation}
where the second matrix ensures that the two eigenvalues of the Majorana
mass matrix $\tilde{M}_{M,0}$ are positive.
Using the Casas-Ibarra parametrization Eq.~(\ref{h}), one can
translate Eq.~(\ref{tildeF}) into
\be\label{condOm}
|\tilde{\O}_{\a2}|\sim \e_h|\tilde{\O}_{\a1}|.
\ee
where $\tilde{\O}\equiv\O U_N^{\dagger}$. From Eq.~(\ref{Omega}) for normal hierarchy,
and choosing the `--' sign and $\xi=1$, one obtains in the limit $|\o|\gg 1$
\begin{equation}
\tilde{\O}(|\o|\gg 1)\sim{1\over \sqrt{2}}\left(
\begin{array}{ccc}
 0  &  0     \\
 -2 \o & -{1\over 2\o} \\
 -2{\rm i} \o & {{\rm i}\over 2\o}
\end{array}\right).
\end{equation}
Note that the case `+' and $\xi=-1$ would have given the same
structure, though with different signs. On the other hand, for the
two other choices, `+' and $\xi=1$, and  `--' and $\xi=-1$, the
columns 1 and 2 would have been exchanged. The case of inverted
hierarchy is obtained by exchanging the rows $2\to 1$ and $3\to 2$.

It is easy to see now that the condition~(\ref{condOm}) is indeed
satisfied when $|\o|\gg 1$, for `--' and $\xi=1$ as well as `+' and
$\xi=-1$, and the breaking parameter $\e_h$ corresponds to \be
\e_h\sim {1\over 4|\o|^2}. \ee

\section{Leptogenesis: non-supersymmetric case}

It is widely known that, in order to generate a baryon asymmetry in the early Universe,
one needs to satisfy the three Sakharov's conditions~\cite{Sakharov:1967dj}.
Baryon number violation is provided by the non-perturbative
sphaleron processes~\cite{Kuzmin:1985mm}. The $C\!P$ symmetry is violated in the decay of the
heavy RH neutrinos~\cite{Covi:1996wh}
\begin{eqnarray}\label{veia}
\ve_{i\a}&\equiv&-{\G(N_i\to \ell_{\alpha}\F)-\G(N_i\to
\bar{\ell}_{\alpha} \F^{\dagger}) \over \sum_{\a}\left[\G(N_i\to
\ell_{\alpha}\F)+\G
(N_i\to \bar{\ell}_{\alpha} \F^{\dagger})\right]}\nonumber\\
&=&\frac{3}{16 \p (h^{\dag}h)_{ii}} \sum_{j\neq i} \left\{ {\rm
Im}\left[h_{\a i}^{\star}
h_{\a j}(h^{\dag}h)_{i j}\right] \frac{\x(x_j/x_i)}{\sqrt{x_j/x_i}}\right.\nonumber \\
&&\hspace{3cm} \left.+ \frac{2}{3(x_j/x_i-1)}{\rm Im}
\left[h_{\a i}^{\star}h_{\a j}(h^{\dag}h)_{j i}\right]\right\} ,
\end{eqnarray}
where $x_i\equiv M_i^2/M_1^2$ and
\be\label{xi}
\xi(x)= {2\over 3}\,x\,
\left[(1+x)\,\ln\left({1+x\over x}\right)-{2-x\over 1-x}\right] \, .
\ee

As for the third condition, namely departure from thermal equilibrium, it can be
conveniently described
by the so-called decay parameter $K_i\equiv \widetilde{\G}_i/H_{T=M_i}$, given by the ratio of the
decay widths to the expansion rate
when the RH neutrinos start to become non-relativistic
at $T=M_i$. In terms of Yukawa couplings, the decay parameters can be written as
\be
K_i={v^2 \over m_{\star} M_i}(h^{\dagger}h)_{ii},
\ee
where $m_{\star}$ is the equilibrium neutrino mass, given by \cite{Buchmuller:2004nz}
\begin{equation}\label{d}
m_{\star} \equiv {16 \pi^{5/2} \sqrt{g_{\star}}\over
3\sqrt{5}}{v^2\over M_{\rm Pl}}\simeq 1.08\times 10^{-3}\,{\rm eV},
\end{equation}
with $M_{\rm Pl}=1.22\times 10^{19}$~GeV and $g_{\star}=g_{\rm
SM}=106.75$.

Notice that, using the Casas-Ibarra parametrization, Eq.~(\ref{h}), the
decay parameters $K_i$ can be expressed as linear combinations
of the neutrino masses \cite{Fujii:2002jw,Buchmuller:2004nz}
\be\label{K}
K_i = \sum_j\,{m_j\over m_{\star}}\,|\O_{ji}^2| \, .
\ee

We will assume in the following that the generation of asymmetry takes
place at temperatures $T\ll 10^9{\rm GeV}$, in which case flavor matters
in the leptogenesis process, and three flavors are distinguished, denoted
$e,\m,\t$ \cite{Nardi:2006fx,Abada:2006fw}.

Important parameters in flavored leptogenesis are the flavored decay
parameters, given by
\be\label{Kflav}
K_{i\a}={v^2 \over m_{\star} M_i}|h_{\a i}|^2.
\ee

For quasi-degenerate RH neutrinos it will prove useful to calculate
the sum of the decay parameters. Using Eqs.~(\ref{K}) and
(\ref{Kflav}) with the matrix $\O$ given in Eq.~(\ref{Omega}) fixing
the sign to '$-$' and $\xi=1$ compatible with the lepton symmetry,
one obtains
\begin{equation}
K_1+K_2={m_2+m_3 \over m_{\star}}(|1-\o^2|+|\o^2|)\geq 56
\end{equation}
and
\begin{eqnarray}
K_{1\a}+K_{2\a}&=&{m_2|U_{\a 2}|^2+m_3|U_{\a 3}|^2 \over m_{\star}}(|1-\o^2|+|\o^2|)\\
&& \hspace{1cm}+ {4\sqrt{m_2 m_3} \over m_{\star}} \,{\rm Im}
\left(U_{\a2}^{\star}U_{\a3}\right) {\rm
Im}\left(\sqrt{1-\o^2}^{\star}\o\right).
\end{eqnarray}
The case of inverted hierarchy is obtained by changing the labels
$3\to 2$ and $2\to 1$. In the limit of large $|\o|$ required by the
symmetry, one obtains
\begin{equation}
K_1+K_2\simeq{2(m_2+m_3) \over m_{\star}}|\o|^2
\end{equation}
and
\begin{equation}
K_{1\a}+K_{2\a}\simeq g_{\a}(m_2,m_3, U_{\a 2},U_{\a3}) |\o|^2,
\end{equation}
which does not depend on the phase of $\o$, and where we defined the
dimensionless quantity
\begin{equation}\label{g}
g_{\a}(m_2,m_3,
U_{\a 2},U_{\a3}) \equiv \left[2(m_2|U_{\a 2}|^2+m_3|U_{\a 3}|^2)-
4\sqrt{m_2 m_3} \,{\rm Im}
\left(U_{\a2}^{\star}U_{\a3}\right)\right]/m_{\star}.
\end{equation}
It is interesting to notice that for a normal hierarchy the small
entry $U_{e3}$ makes possible a cancellation in $K_{1e}+K_{2e}$ such
that it can be much smaller than $K_1+K_2\gg 56$; the other two
flavors cannot be smaller than roughly $(K_1+K_2)/6$. On the other
hand, when the hierarchy is inverted, a cancellation can occur in
all flavors.

For quasi-degenerate RH neutrinos and with a strong washout in each
flavor, the solution to the flavored Boltzmann equations can be
written in the form~\cite{Blanchet:2006be,Blanchet:2006dq}
\begin{equation}\label{Nf}
N_{B-L}^{\rm f}=\sum_{\alpha}N_{\D_{\a}}^{\rm
f}=\sum_{\alpha}(\ve_{1\a}+\ve_{2\a}) \,\kappa(K_{1\a}+K_{2\a}),
\end{equation}
where $\kappa$ is the so-called efficiency factor, given by
\cite{Buchmuller:2004nz,Blanchet:2006be}
\begin{equation}\label{kappa}
\k(K) \equiv {2\over K\,z_B(K)}\, \left(1-{\rm e}^{-{K\,z_B(K)\over
2}}\right) \, ,
\end{equation}
where
\begin{equation}
z_{B}(K) \simeq 2+4\,K^{0.13}\,{\rm
e}^{-{2.5\over K}} \, .
\end{equation}
Note that in the strong washout regime, for $K\gg 1$, the following
approximation holds~\cite{DiBari:2004en,Giudice:2003jh}:
\begin{equation}
\label{kappaapprox} \kappa(K)\simeq 0.5/K^{1.16}.
\end{equation}

As for the $C\!P$ asymmetries, in the limit $M_1\simeq M_2$ and for the particular
$\O$ matrix in Eq.~(\ref{Omega})\footnote{The
following relations hold for this model: $(h^{\dagger}h)_{13}=(h^{\dagger}h)_{31}=0,\quad
(h^{\dagger}h)_{23}=(h^{\dagger}h)_{32}=0$}, they are given by
\begin{eqnarray}
\ve_{1\alpha}&\simeq& \frac{3}{16 \p (h^{\dag}h)_{11}} \frac{1}{3\d_2} \left({\rm Im}
\left[h_{\a 1}^{\star} h_{\a 2}(h^{\dag}h)_{1 2}\right]+{\rm Im}\left[h_{\a 1}^{\star} h_{\a 2}
(h^{\dag}h)_{2 1}\right]\right),\\
\ve_{2\alpha}&\simeq& \frac{3}{16 \p (h^{\dag}h)_{22}} \frac{1}{3\d_2} \left({\rm Im}
\left[h_{\a 1}^{\star} h_{\a 2}(h^{\dag}h)_{1 2}\right]+{\rm Im}\left[h_{\a 1}^{\star} h_{\a 2}
(h^{\dag}h)_{2 1}\right]\right),
\end{eqnarray}
where $\d_2\equiv M_2/M_1 -1$. Using the Casas-Ibarra
parametrization Eq.~(\ref{h}) the flavored $C\!P$ asymmetries are
given by
\begin{eqnarray}
\ve_{1\alpha}&\simeq& \frac{M_2}{16 \p v^2} {1\over \delta_2} {1\over m_2|1-\o^2|
+m_3|\o^2|}\times \nonumber\\
&& \quad \left\{ \left( m_2^2|U_{\a 2}|^2 -m_3^2|U_{\a 3}|^2 +m_2 m_3 (|U_{\a 3}|^2-
|U_{\a 2}|^2)\right)\,{\rm Im} (\o^2)\right. \\
&& \left.\qquad - 2 (m_3-m_2)\sqrt{m_2m_3}\,{\rm Im} (U_{\a
2}^*U_{\a 3})
\,{\rm Re} \left(\o \sqrt{1-\o^2}\right)\right\}\nonumber \\
\ve_{2\alpha}&\simeq& \frac{M_1}{16 \p v^2} {1\over \delta_2} {1\over m_2|\o^2|
+m_3|1-\o^2|}\times \nonumber\\
&& \quad \left\{ \left( m_2^2|U_{\a 2}|^2 -m_3^2|U_{\a 3}|^2 +m_2 m_3 (|U_{\a 3}|^2-
|U_{\a 2}|^2)\right)\,{\rm Im} (\o^2)\right.\\
&& \left.\qquad - 2 (m_3-m_2)\sqrt{m_2m_3}\,{\rm Im} (U_{\a
2}^*U_{\a 3}) \,{\rm Re}
\left(\o\sqrt{1-\o^2}\right)\right\}\nonumber.
\end{eqnarray}
In the limit of large $|\o|$ and defining $\o\equiv |\o|(\cos
\q_{\o}+{\rm i}\sin\q_{\o})$, one obtains
\begin{equation}\label{epslargeo}
\ve_{1\alpha}\simeq \ve_{2\a}\simeq \frac{M m_{\star}}{16 \p v^2}
{1\over \delta_2} f_{\a}(m_2,m_3, U_{\a 2},U_{\a3})\,\sin 2\q_{\o},
\end{equation}
where we defined another dimensionless quantity, \bea\label{f}
f_{\a}(m_2,m_3, U_{\a 2},U_{\a3})&\equiv& {g_{\a}\over
2}{m_2-m_3\over m_2+m_3}. \eea It is interesting to notice first
that the $C\!P$ asymmetry is independent of $|\o|$, and second that
the high-energy contribution is a necessary ingredient to have a
non-zero flavored $C\!P$ asymmetry; in other words, $\o$ has to be
complex.  Hence, leptogenesis from exclusively low-energy
phases~\cite{Blanchet:2006be,Pascoli:2006ie,Pascoli:2006ci,Branco:2006ce,Molinaro:2007uv,Anisimov:2007mw}
is not possible in our model.

We have pointed out earlier that $K_{1\a}+K_{2\a}$ can be much
smaller than $K_1+K_2$ in certain situations, which could lead to a
big enhancement of the predicted baryon asymmetry compared to an
unflavored calculation. However, it turns out that in all such cases
the flavored $C\!P$ asymmetry is suppressed as well, so that the
final effects are never much larger than the typical enhancement of
a factor three in the three-flavor regime, as we will show more
precisely below. Since the lepton number symmetry implies $|\o| \gg
1$, the asymmetry will be typically produced in the strong washout
regime, with no dependence on the initial number of RH neutrinos and
on any previously generated asymmetry. With an account of flavor
effects, the second feature is only rigourously possible in the
three-flavor regime, where the washout occurs in all directions in
flavor space~\cite{Engelhard:2006yg}. The only exception will be in
some very marginal regions of the parameter space where huge flavor
effects imply $K_{1\a}+K_{2\a}\lesssim 3$, even though $K_1+K_2\gg
56$, so that the strong washout and independence of the initial
conditions are no longer guaranteed. We will come back to this point
at the end of the section.

Since the final asymmetry will be almost exclusively produced in the
strong washout, using Eqs.~(\ref{Nf}), (\ref{kappaapprox}) and
(\ref{epslargeo}), one obtains the simple expression
\begin{equation}\label{resultanal2}
N_{B-L}^{\rm f}\simeq {M m_{\star}\over 16 \p v^2} {\sin
2\q_{\o}\over \d_2 |\o|^{2.32}} \sum_{\a} {f_{\a}\over
g_{\a}^{1.16}},
\end{equation}
which depends on the fundamental
quantities $M$, $\d_2$, $|\o|$ and $\q_{\o}$ in a very simple way.
In terms of the breaking parameters $\e_M$ and $\e_h$ introduced in
the last section, one has
\begin{equation}
\label{resultanal} N_{B-L}^{\rm f}\simeq {M m_{\star}\over 3 \p v^2}
{\e_h^{1.16}\over \e_M }\sin 2\q_{\o} \sum_{\a} {f_{\a}\over
g_{\a}^{1.16}}.
\end{equation}
Interestingly, all low-energy parameters (neutrino masses, PMNS
matrix elements) appear exclusively in the factor $\sum_{\a}
f_{\a}/g_{\a}^{1.16}$, and are thus decoupled from the high-energy
parameters ($M$, $\d_2$, $|\o|$ and $\q_{\o}$). Actually, flavor
effects altogether only appear in this factor, and we can thus
easily estimate the maximal difference with an unflavored
calculation, where we would have instead a factor
\begin{equation}
{\sum_{\a}f_{\a}\over \sum_{\a}g_{\a}^{1.16}}={m_2-m_3\over
2^{1.16}(m_2+m_3)^{1.16}}
\simeq \left\{\begin{array}{ll} 0.16 &\textrm{normal hierarchy}\\
1.7\times 10^{-3}& \textrm{inverted hierarchy}\end{array}\right. .
\end{equation}
On the other hand, one can find numerically the maximal factor in
the flavored calculation:
\begin{equation}\label{maxfoverg} {\rm
max}\left(\sum_{\a}{f_{\a}\over g_{\a}^{1.16}}\right)
\simeq \left\{\begin{array}{ll} 0.87 &\textrm{normal hierarchy}\\
9\times 10^{-3}& \textrm{inverted hierarchy}\end{array}\right. ,
\end{equation}
i.e. maximal flavor effects lead to a factor 5--6 enhancement of the
final asymmetry both for normal and for inverted hierarchy. As we
said earlier, larger effects are not possible here.

Finally, assuming a standard thermal history and accounting for the
sphaleron conversion coefficient $a_{\rm sph}\sim 1/3$, the
baryon-to-photon ratio can be calculated as \be\label{etaB}
\eta_B=a_{\rm sph}\,{N_{B-L}^{\rm f}\over N_{\gamma}^{\rm rec}}
\simeq 0.96\times 10^{-2}\,N_{B-L}^{\rm f} \, , \ee to be compared
with the measured value \cite{Komatsu:2008hk} \be\label{etaBobs}
\eta_B^{\rm CMB} = (6.2 \pm 0.15)\times 10^{-10} \, . \ee

It can be useful to make a numerical estimation of $\eta_B$ using
Eq.~(\ref{resultanal}), with $\q_{\o}=-\p/4$, and
$\sum_{\a}f_{\a}/g_{\a}^{1.16}=0.7$:
\begin{equation}
\eta_B\sim 3.6\times 10^{-10} \left({\e_h^{1.16}\over \e_M}\right)
\left({M_1\over 10^{10}~{\rm GeV}}\right).
\end{equation}
This result tells us that a hierarchy in the breaking parameters
$\e_h\gg \e_M$ is needed if one wants to relax the scale of
leptogenesis. For instance, with the breaking parameters
$\e_h=2.5\times 10^{-3}$ and $\e_M=4\times 10^{-8}$, one obtains
\begin{equation}\label{estimation}
\eta_B \sim 6\times 10^{-10}\left({\e_h\over 2.5\times
10^{-3}}\right)^{1.16} \left({4\times 10^{-8}\over
\e_M}\right)\left({M_1\over 10^6~{\rm GeV}}\right).
\end{equation}
It can be easily seen from Eq.~(\ref{resultanal}) that the scale of
the RH neutrinos $M$ can be lowered if we decrease by the same
factor $\e_M$. Hence, with $\e_M=4\times 10^{-11}$, we can reach the
TeV scale for the heavy neutrinos, making them at least in principle
accessible at the LHC.

The $C\!P$ asymmetry is enhanced in the degenerate limit inversely
proportionally to $\d_2$, as can be seen from Eq.~(\ref{epslargeo}).
However, this effect is not unlimited. There is a maximal
enhancement which leads to resonant
leptogenesis~\cite{Pilaftsis:2003gt,Pilaftsis:2005rv}. The condition
to be on the resonance is given by
$M_j-M_i=\G_j/2$~\cite{Pilaftsis:1997jf}, where $\G_j$ is the decay
width of $N_j$. In our model and for the case of normal hierarchy,
this condition can be conveniently translated into
\begin{equation}\label{condres}
\d^{\rm res}= d {m_{\rm atm} M |\o|^2\over 16 \p v^2}\simeq 3\times
10^{-11}\,d\, |\o|^2 \left({M\over 10^6~{\rm GeV}}\right),
\end{equation}
where we introduced the parameter $d$ as in
\cite{Anisimov:2007mw} to account for a controversy in the
literature about whether it is allowed or not to reach the
resonance~\cite{Pilaftsis:1997jf,Anisimov:2005hr}. In the following,
we will use conservatively $d=5$, so that the validity of
Eq.~(\ref{veia}) is ensured. Note that in the case of inverted
hierarchy $\d^{\rm res}$ occurs at a value twice as large. In the
following, we will assume that $\e_M\geq \d^{\rm res}$.

Plugging the condition~(\ref{condres}) in Eq.~(\ref{resultanal2}),
we find for normal hierarchy
\begin{equation}
\eta_{B}\leq 0.96\times 10^{-2}\,{m_{\star} \over d m_{\rm atm}
|\o|^{4.32}} {\rm max}\left(\sum_{\a} {f_{\a}\over
g_{\a}^{1.16}}\right)\simeq {3.7\times 10^{-5}\over |\o|^{4.32}},
\end{equation}
where we used Eq.~(\ref{maxfoverg}) in the second step. Since we
want to be consistent with the $3\s$ range of Eq.~(\ref{etaBobs}),
we find that
\begin{equation}\label{resnormal}
|\o|^2\lesssim 168 \quad \Rightarrow \quad \e_h\gtrsim 1.3\times
10^{-3}.
\end{equation}
In the case of inverted hierarchy, we find
\begin{equation}
\eta_{B}\leq 0.96\times 10^{-2}\,{m_{\star} \over d m_{\rm atm}
|\o|^{4.32}} {\rm max}\left( \sum_{\a}{f_{\a}\over
g_{\a}^{1.16}}\right)\simeq {1.8\times 10^{-7}\over |\o|^{4.32}}.
\end{equation}
This implies
\begin{equation}\label{resinverted}
 |\o|^2\lesssim
14\quad \Rightarrow \quad \e_h\gtrsim 1.8\times 10^{-2}.
\end{equation}
Remember that the lepton number symmetry implied $|\o|\gg 1$, so
that the allowed range in the case of inverted hierarchy is quite
constrained! Note also that the case of inverted hierarchy is more
constrained than normal hierarchy due to the smallness of the factor
$\sum_{\a}f_{\a}/g_{\a}^{1.16}$, never larger than $10^{-2}$. This
behavior is different from the 2RHN model in the hierarchical limit
$M_1\ll M_2$, where huge flavor effects make possible that inverted
hierarchy yields almost the same bounds as normal
hierarchy~\cite{Blanchet:2008pw}, contrary to the unflavored
result~\cite{Petcov:2005jh,Blanchet:2006dq}.

The results of Eqs.~(\ref{resnormal}) and (\ref{resinverted}) show
explicitly that, in our simple model with two RH neutrinos or when
$N_3$ is decoupled, successful leptogenesis is only possible for
relatively large values of the breaking parameter $\e_h$, especially
in the case of inverted hierarchy. This is particularly interesting
in view of the possible observation of RH neutrinos at future
colliders, as recently investigated in a number of
papers~\cite{delAguila:2005mf,Han:2006ip,delAguila:2007em,Kersten:2007vk}.
To have even a small chance of observing RH neutrinos,  some of the
active-sterile mixing angles $V_{\a i}=(M_D/M_M)_{\a i}$ should not
be much smaller than 0.01. In our case, with the constraint from
successful leptogenesis, we have at most $V\sim 10^{-5}$ for
$M=250$~GeV, which is much too small. Lepton-flavor-violating
signals in the non-supersymmetric case under discussion are expected
to be very suppressed for the same reason.

\begin{figure}
\includegraphics[width=0.35\textwidth,angle=-90]{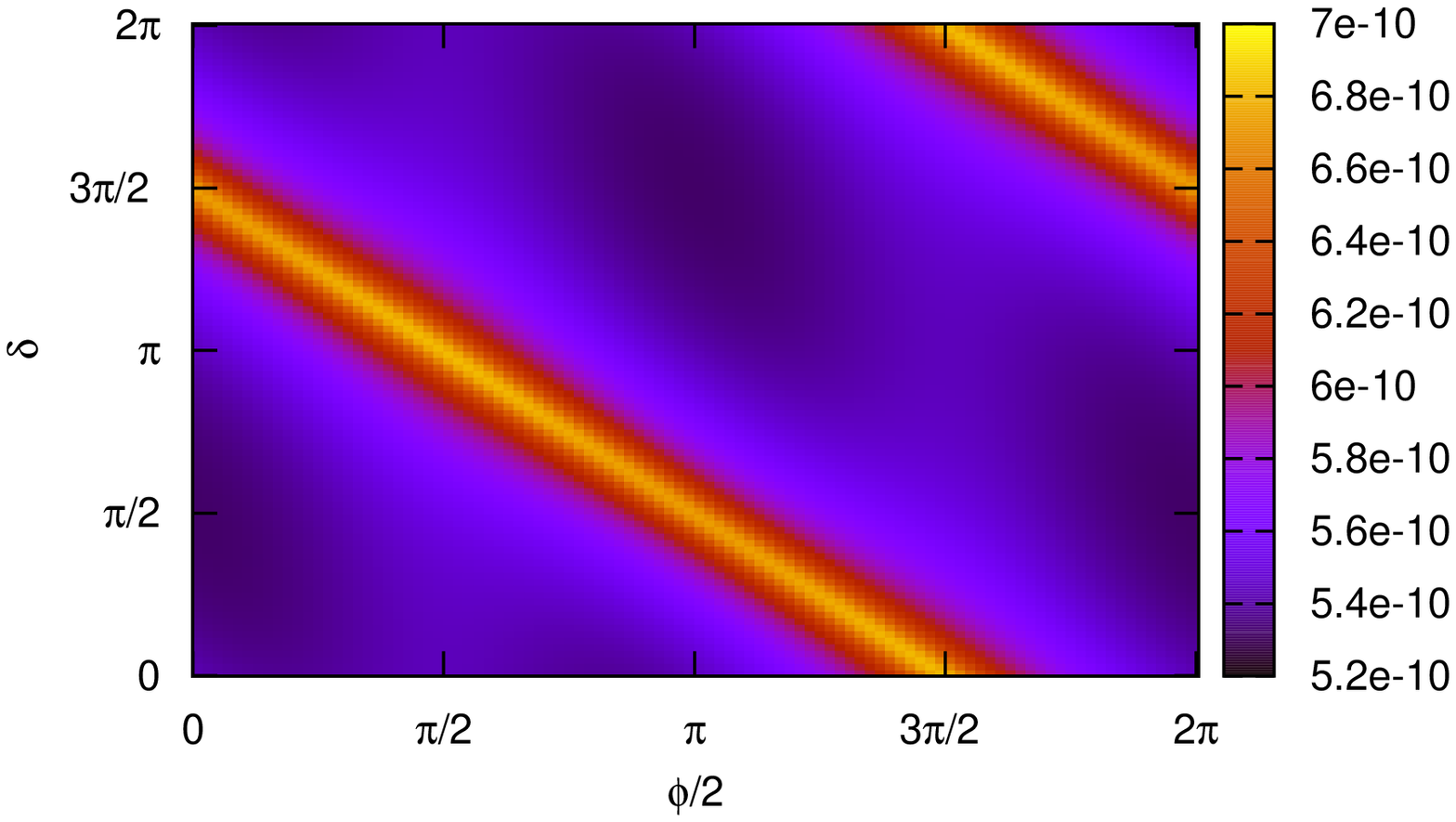}
\hspace{0.1cm}
\includegraphics[width=0.35\textwidth,angle=-90]{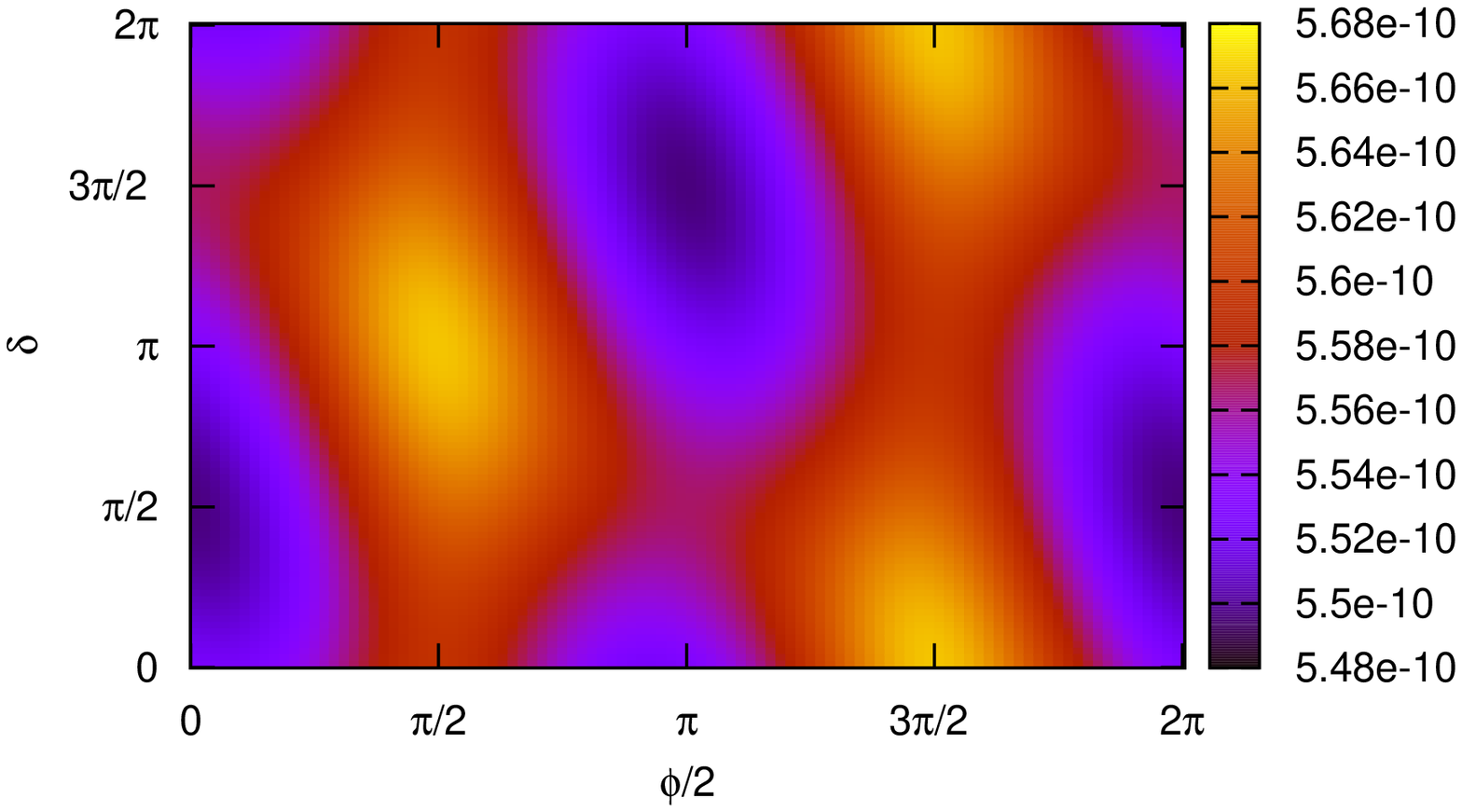}
\caption{Dependence of the baryon-to-photon ratio $\eta_B$ on the
Dirac phase $\d$ and the Majorana phase $\phi$. We display the case
of normal hierarchy, with $\q_{\o}=-\p/4$, $M_1=10^{6}~{\rm GeV}$,
$|\o|^2=100$ (implying $\e_h\sim 2.5\times 10^{-3}$),
$\d_{2}=4\times 10^{-8}\sim \e_M$. Left panel: $\sin \q_{13}= 0.2$;
Right panel: $\sin \q_{13}= 0.02$.} \label{fig:correlnormal}
\end{figure}

\begin{figure}
\includegraphics[width=0.35\textwidth,angle=-90]{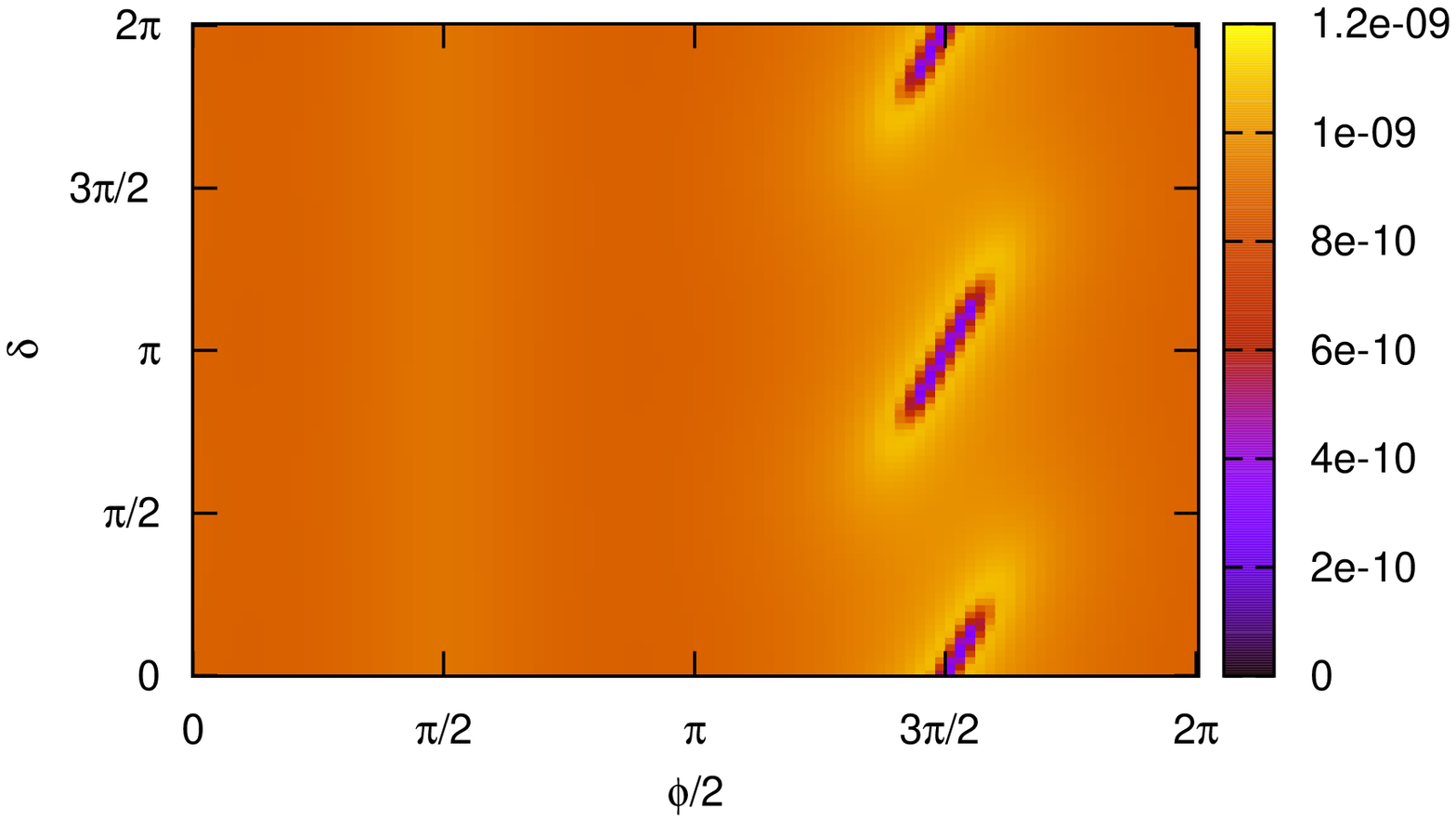}
\hspace{0.1cm}
\includegraphics[width=0.35\textwidth,angle=-90]{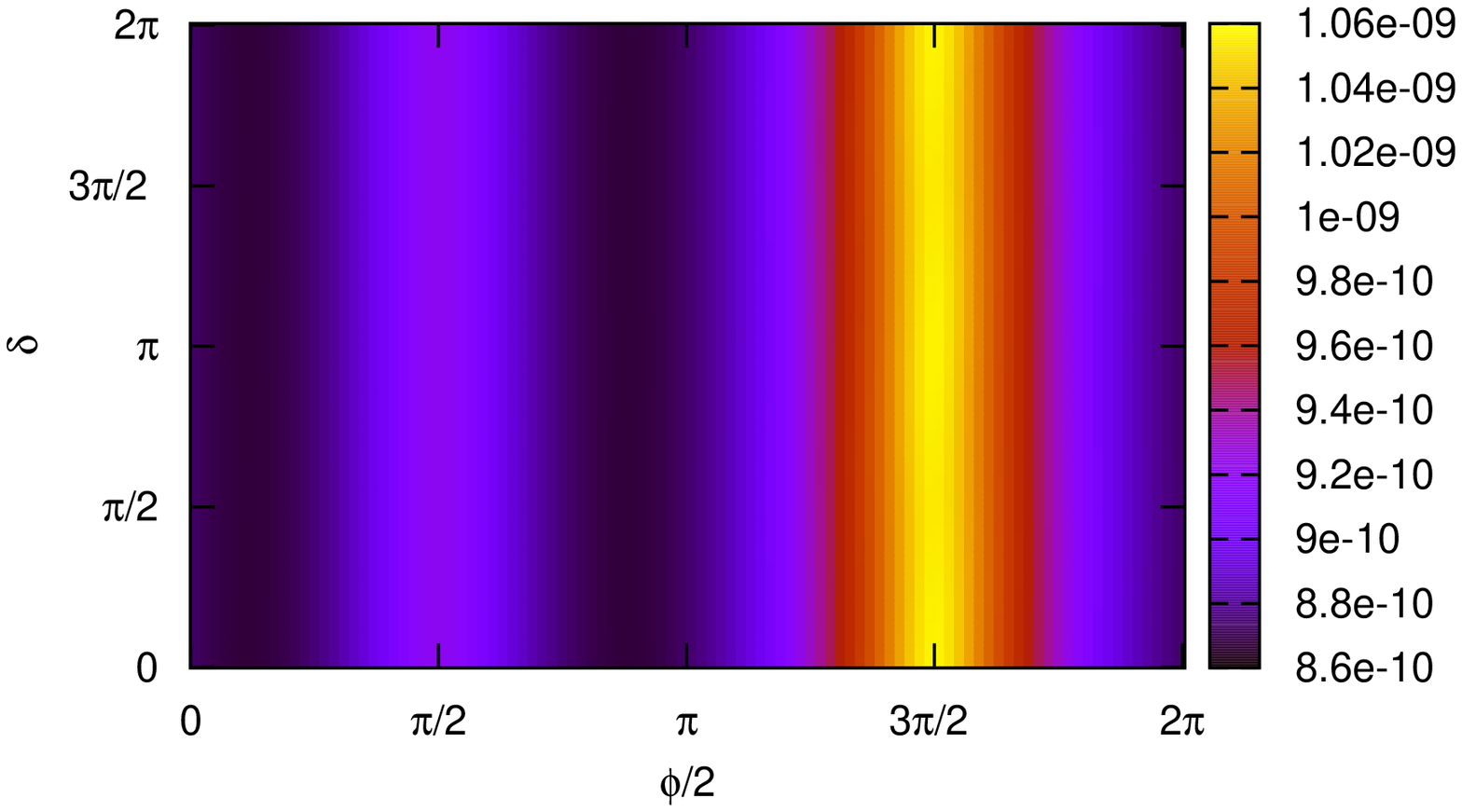}
\caption{Same as Fig.~\ref{fig:correlnormal}, but for inverted
hierarchy, with $|\o|^2=10$ (implying $\e_h\sim 2.5\times 10^{-2}$),
$\d_{2}=3.3\times 10^{-9}\sim \e_M$.} \label{fig:correlinverted}
\end{figure}

Finally, we would like to present examples compatible with the
constraints~(\ref{resnormal}) and (\ref{resinverted}) which show the
explicit dependence of $\eta_B$ on the angle $\q_{13}$ and on the
PMNS phases $\d$ and $\phi$, which only appear in the factor
$\sum_{\a} f_{\a}/g_{\a}^{1.16}$, as already mentioned. We present
in Fig.~\ref{fig:correlnormal} two such examples, for a normal
hierarchy of light neutrinos and two choices of $\sin \q_{13}$, 0.2
and 0.02. Note that the results are compatible with our rough
estimation in Eq.~(\ref{estimation}). In
Fig.~\ref{fig:correlinverted} we display the case of inverted
hierarchy. One notices from the figures that the two
$C\!P$-violating phases in the PMNS matrix only yield small
corrections to the predicted baryon asymmetry. When the hierarchy of
light neutrinos is normal and for the maximal allowed value
$\sin\q_{13}=0.2$, these phases can change the final asymmetry by
40\%. When the hierarchy is inverted, the effect can be more than a
factor three, but only in a very restricted region of the parameter
space.

Before concluding the section, one comment concerning
Figs.~\ref{fig:correlnormal} and ~\ref{fig:correlinverted} is in
order. We found marginal regions where $K_{1\a}+K_{2\a}\lesssim 3$
for some flavor $\a$, implying that the strong washout regime and
thus Eq.~(\ref{kappaapprox}) no longer hold. In these regions of
dependence on the initial conditions, we used the more general
Eq.~(\ref{kappa}), which is also valid for $K\lesssim 3$ in the case
of thermal initial $N_1$ and $N_2$ abundances. In practice, these
regions of huge flavor effects on the washout are found very close
to the ones where the asymmetry is maximal, and they can be seen in
the left panels of Figs.~\ref{fig:correlnormal} and
~\ref{fig:correlinverted}. Note that a very small $\sin \theta_{13}$
as in the right panels forbids such huge suppressions of the
washout. More specifically, in the left panel of
Fig.~\ref{fig:correlnormal}, $K_{1\alpha}+K_{2\a}$ can be as low as
$\sim 1$ in the middle of the diagonal strips, implying some
dependence on the initial conditions. In the left panel of
Fig.~\ref{fig:correlinverted}, $K_{1\alpha}+K_{2\a}$ can be as low
as $\sim 0.7$ in the darker regions (red to black), so the
dependence on the initial conditions is even larger there. But apart
from these marginal regions, the strong washout holds, and one can
safely use Eq.~(\ref{resultanal}).

\section{Leptogenesis: supersymmetric case}

We turn now to the supersymmetric version of the model. One can
introduce four different $C\!P$ asymmetries, which by supersymmetry
are all equal:
\begin{eqnarray}
\tilde{\ve}_{i\a}&\equiv&-{\G(N_i\to \ell_{\alpha} H_u)-\G(N_i\to
\bar{\ell}_{\alpha} H_u^{\dagger}) \over \sum_{\a}\left[\G(N_i\to
\ell_{\alpha}H_u)+\G (N_i\to \bar{\ell}_{\alpha}
H_u^{\dagger})\right]}=-{\G(N_i\to
\tilde{\ell}_{\alpha}\tilde{H}_u-\G(N_i\to
\tilde{\ell}^{\dagger}_{\alpha} \bar{\tilde{H}}_u)) \over
\sum_{\a}\left[\G(N_i\to \tilde{\ell}_{\alpha} \tilde{H}_u)+\G
(N_i\to \tilde{\ell}^{\dagger}_{\alpha} \bar{\tilde{H}}_u)\right]} \nonumber\\
&=&-{\G(\tilde{N}_i\to
\ell_{\alpha}\tilde{H}_u)-\G(\tilde{N}^{\star}_i\to
\bar{\ell}_{\alpha} \bar{\tilde{H}}_u) \over
\sum_{\a}\left[\G(\tilde{N}_i\to
\ell_{\alpha}\tilde{H}_u)+\G(\tilde{N}^{\star}_i\to
\bar{\ell}_{\alpha} \bar{\tilde{H}}_u)\right]}=-{\G(\tilde{N}_i\to
\tilde{\ell}_{\alpha}H_u)-\G(\tilde{N}^{\star}_i\to
\tilde{\ell}^{\dagger}_{\alpha} H_u^{\dagger}) \over
\sum_{\a}\left[\G(N_i\to \tilde{\ell}_{\alpha} H_u)+\G
(\tilde{N}^{\star}_i\to \tilde{\ell}^{\dagger}_{\alpha}
H_u^{\dagger})\right]}\nonumber \, ,
\end{eqnarray}
where $\tilde{\ell}$, $\tilde{H}_u$ and $\tilde{N}$ denote sleptons,
higgsinos and RH sneutrinos, respectively. The $C\!P$ asymmetries
$\tilde{\ve}_{i\a}$ were calculated in~\cite{Covi:1996wh} to be
\begin{equation}
\tilde{\ve}_{i\a}=\frac{1}{8 \p (h^{\dag}h)_{ii}} \sum_{j\neq i}
\left\{ {\rm Im}\left[h_{\a i}^{\star} h_{\a j}(h^{\dag}h)_{i
j}\right] g(x_j/x_i)+ \frac{2}{(x_j/x_i-1)}{\rm Im} \left[h_{\a
i}^{\star}h_{\a j}(h^{\dag}h)_{j i}\right]\right\} ,
\end{equation}
where $x_i\equiv M_i^2/M_1^2$ and \be\label{xi} g(x)= \sqrt{x}
\left[{2\over x-1}+\ln\left({1+x\over x}\right)\right] \, . \ee

In the limit $M_1\simeq M_2$ we are interested in and for the
particular $\O$ matrix in Eq.~(\ref{Omega}), it can be easily
obtained that
\begin{eqnarray}
\tilde{\ve}_{1\alpha}&\simeq& \frac{1}{8 \p (h^{\dag}h)_{11}}
\frac{1}{\d_2} \left({\rm Im} \left[h_{\a 1}^{\star} h_{\a
2}(h^{\dag}h)_{1 2}\right]+{\rm Im}\left[h_{\a 1}^{\star} h_{\a 2}
(h^{\dag}h)_{2 1}\right]\right),\\
\tilde{\ve}_{2\alpha}&\simeq& \frac{1}{8 \p (h^{\dag}h)_{22}}
\frac{1}{\d_2} \left({\rm Im} \left[h_{\a 1}^{\star} h_{\a
2}(h^{\dag}h)_{1 2}\right]+{\rm Im}\left[h_{\a 1}^{\star} h_{\a 2}
(h^{\dag}h)_{2 1}\right]\right),
\end{eqnarray}
where we recall that $\d_2\equiv M_2/M_1 -1$. So the $C\!P$
asymmetries are a factor of two larger than in the
non-supersymmetric case, just like in the hierarchical limit $M_2\gg
M_1$.

Let us see how the efficiency factor is affected by supersymmetry.
Assuming Maxwell-Boltzmann distributions, RH neutrinos and RH
sneutrinos follow exactly the same evolution. The end result is that
the source term in the Boltzmann equation for the $B/3-L_{\alpha}$
asymmetry (see e.g. \cite{Antusch:2006cw,Abada:2008gs}) is a factor
4 larger than in the non-SUSY case, whereas the washout term is a
factor 2 larger. The reason is that, first, the $C\!P$ asymmetry is
a factor 2 larger, and second there are twice as many decay modes as
in the non-SUSY case\footnote{The fact that both the decays of RH
neutrinos and RH sneutrinos contribute is approximately balanced by
the larger dilution factor in the SUSY case.}. The latter of course
affects both source and washout terms. Finally, the equilibrium
neutrino mass in Eq.~(\ref{d}) is numerically different since
$g_{\star}=g_{\rm MSSM}=228.75$, and $v_u=v\, \sin \beta$, implying
\begin{equation}
m_{\star}^{\rm MSSM}\simeq (1.56\times 10^{-3}~{\rm eV})\sin^2\beta.
\end{equation}

Altogether, multiplying the $C\!P$ asymmetry by a factor 2, and
replacing $K_{1\alpha}+K_{2\alpha} \to 2(K_{1\alpha}+K_{2\alpha})$,
the semi-analytical expression for the final asymmetry becomes [cf.
Eq.~(\ref{resultanal})]
\begin{equation}\label{resultanalSUSY}
N_{B-L}^{\rm f}\simeq {2 M m_{\star}^{\rm MSSM}\over 3 \p v_u^2}
{\e_h^{1.16}\over \e_M }\sin 2\q_{\o} \sum_{\a} {f'_{\a}\over (2
g'_{\a})^{1.16}},
\end{equation}
where $f'_{\alpha}$ and $g'_{\alpha}$ are the functions defined in
Eqs.~(\ref{f}) and (\ref{g}), but with the replacement $m_{\star}\to
m_{\star}^{\rm MSSM}$. Note that the dependence on $\sin \beta$ in
this expression is extremely mild, $N_{B-L}^{\rm f}\propto\sin
\beta^{0.16}$.

The extension of the results presented in the last section is then
straightforward. From the maximal enhancement of the $C\!P$
asymmetry in the degenerate limit [cf. Eq.~(\ref{condres})], we
obtain slightly modified bounds on the breaking parameters,
\begin{equation}\label{resnormalSUSY}
|\o|^2\lesssim 160 \quad \Rightarrow \quad \e_h\gtrsim 1.5\times
10^{-3},
\end{equation}
in the case of normal hierarchy, and
\begin{equation}\label{resinvertedSUSY}
 |\o|^2\lesssim
14\quad \Rightarrow \quad \e_h\gtrsim 1.8\times 10^{-2}
\end{equation}
in the case of inverted hierarchy.

Since the limits on the breaking parameters are very similar to the
non-SUSY case, the active-sterile mixing angles are necessarily
small here as well. Hence, the conclusion about the incompatibility
of leptogenesis with the observation of RH neutrinos at colliders is
still valid. In the case of lepton-flavor violation, the discussion
is somewhat different since new diagrams contribute that are not
suppressed by the small mixing between active and sterile
neutrinos~\cite{Borzumati:1986qx}. For example, the rate of $\mu\to
e\gamma$ will actually depend on the combination
$(hh^{\dagger})_{e\mu}$. However, for the range of RH neutrino
masses we are interested in to avoid the gravitino problem, i.e.
$M<10^6$~GeV,  the Yukawa couplings are too small ($h\lesssim
10^{-3}$) to give an observable signal.

\section{Summary and conclusion}

We have studied the mechanism of leptogenesis in the presence of a
slightly broken lepton number symmetry. Two almost degenerate
right-handed neutrinos result from the symmetry, with the small
breaking parameter $\e_M$ essentially describing the mass splitting,
and $\e_h$ fixing the size of the (inverse of the) washout. Two
well-known consequences follow: First, the scale of leptogenesis can
be as low as the electroweak scale. Second, the baryon asymmetry
predicted through leptogenesis is independent of the initial
conditions, i.e. both the initial number of RH neutrinos and any
previously generated asymmetry.

With only two heavy neutrinos, which is equivalent to the $N_3$
decoupling limit, the model contains few unknown parameters in
addition to the two breaking parameters: one RH neutrino mass, one
`high-energy' phase, the angle $\q_{13}$ and two $C\!P$-violating
phases in the PMNS matrix. The relatively low number of parameters
has allowed us to have a perfect handle on the problem, and we have
been able to obtain semi-analytical formulae for the final
asymmetry, Eqs.~(\ref{resultanal}) and (\ref{resultanalSUSY}) in the
non-SUSY and SUSY case, respectively, which disclose a simple
dependence on each one of these parameters. In particular, we have
studied in detail the role of the PMNS phases and $\q_{13}$ (see
Figs.~\ref{fig:correlnormal} and \ref{fig:correlinverted}).
Interestingly, the high-energy phase is required to be non-zero for
successful leptogenesis, implying that leptogenesis from exclusively
low-energy $C\!P$-violation is not viable in this context.

Finally, we derived from successful leptogenesis and the maximal
enhancement of the $C\!P$ asymmetry in the resonant limit that the
breaking parameter $\e_h$ must be relatively large, $\e_h\gtrsim
10^{-3}$ for normal and $10^{-2}$ for inverted hierarchy (both in
SUSY and non-SUSY cases). As a consequence, leptogenesis is not
compatible with the observation of RH neutrinos at future colliders
and with a sizable lepton-flavor violation signal. The other
breaking parameter, $\e_M$, can be much smaller, and actually needs
to be so in order to have low-scale leptogenesis.

\paragraph{Acknowledgments}

It is a pleasure to thank Pasquale Di Bari, J\"orn Kersten, and
Georg Raffelt for useful discussions. S. B. is grateful to the
Max-Planck-Institute for Physics in Munich where most of this work
was done.

\bibliographystyle{JHEP-2}
\providecommand{\href}[2]{#2}\begingroup\raggedright\endgroup

\end{document}